# An Integrated Simulation Tool for Dark Current Radiation Effects Using ACE3P and Geant4*

*Lixin Ge, Zenghai Li, Cho-Kuen Ng, Liling Xiao*
*SLAC, Stanford, CA 94025, USA*

*Hiroyasu Ego, Yoshinori Enomoto, Hiroshi Iwase, Yu Morikawa, Takashi Yoshimoto*
*KEK , 1-1 Oho, Tsukuba, Ibaraki, Japan*

## ABSTRACT

A simulation workflow is under development to interface particle data transfer and matching of geometry between the electromagnetic (EM) cavity simulation code ACE3P and radiation code Geant4. The target is to simulate dark current (DC) radiation effects for the KEK 56-cell S-band accelerating structure using ACE3P and Geant4, and benchmark against KEK experiment data. As a first step, ACE3P DC simulations using a 7-cell structure have been performed by first calculating the operating mode in the structure and then tracking field-emitted electrons under the influence of the EM fields of the mode. The ACE3P simulation results agree well with the EM software CST for an accelerating gradient of 21.8 MV/m. The reader/writer I/O in ACE3P and the transfer of particle data from Track3P to Geant4 for DC radiation effects studies have been implemented. The simulation workflow between the two codes will be demonstrated with the goal of performing large-scale simulations for the KEK 56-cell structure. In addition to modeling DC effects in linacs, the integrated simulation workflow will be applicable to studying positron source and capture structure for future lepton colliders.

## 1. INTRODUCTION

Accelerator cavities operating at high gradients are subject to damage from high energy electrons hitting the surface of the cavity wall. These electrons are believed to originate from certain locations of the cavity surface due to field emission, and then accelerated under the rf field of the cavity before impacting the cavity wall. These electrons will interact with the cavity wall made of certain materials and produce electromagnetic radiation that can affect the performance of the accelerator.

**Talk presented at the International Workshop on Future Linear Colliders (LCWS 2023), 15-19 May 2023. C23-05-15.3.**
* Supported by HEP US - Japan Science and Technology Cooperation Program (2022-2025)

The present approach of radiation calculation involves separate simulations such as using ACE3P [1-5] and Geant4 [6-8], or ACE3P and FLUKA [9,10]. In this paper, we aim to streamline the code integration process using Geant4 instead of FLUKA because the latter is proprietary and code developers generally have no access to the source codes. Geant4, on the other hand, is open source and a widely accepted radiation modeling package.

The current integrated simulation process for large accelerator systems presents several challenges, such as long simulation times, the need for multiple experts from different physics domains, and difficulties in sharing and replicating simulations within the community. It is of utmost importance to provide an easy-to-use and streamlined tool for the accelerator community to better utilize the resources.

Supported by HEP US - Japan Science and Technology Cooperation Program, a standalone software package for modeling particle-matter interactions radiation effects in accelerator cavity is under development by integrating ACE3P for modeling of accelerator structures and Geant4 for calculation of interactions of particles with matter. The work is to integrate the separate calculations into a single simulation workflow from start to end without the need for individual scientists performing different tasks and communicating with each other.

This paper will describe the methods used in the integration package in Section 2: 2.1 is a schematic of the simulation workflow for the integrated software; 2.2 is the basic information of two integrated software ACE3P and Geant4; 2.3 shows how to deal with geometry CAD model in both ACE3P and Geant4; 2.4 describes a particle data transfer capability between ACE3P and Geant4 based on the standardized openPMD format. Preliminary dark current application using the developed simulation tool is presented in Section 3, followed by conclusion and future work in Section 4.

## 2. METHODS FOR ACE3P-GEANT4 INTEGRATION

### 2.1 Simulation Workflow

A schematic of the simulation workflow for the integrated software is illustrated in Fig. 1. It starts with the construction of the geometrical models as two separate computation domains for ACE3P and Geant4 simulations. An integrated code driver first assigns the problem type to determine which code initiates a simulation: ACE3P for dark current and Geant4 for positron source simulations, respectively. Particles hitting the geometrical interface between the two computational domains will be collected and transferred to the other code for its respective physics simulation. Finally, the particle data and radiation distribution will be output to files for visualization and postprocessing.

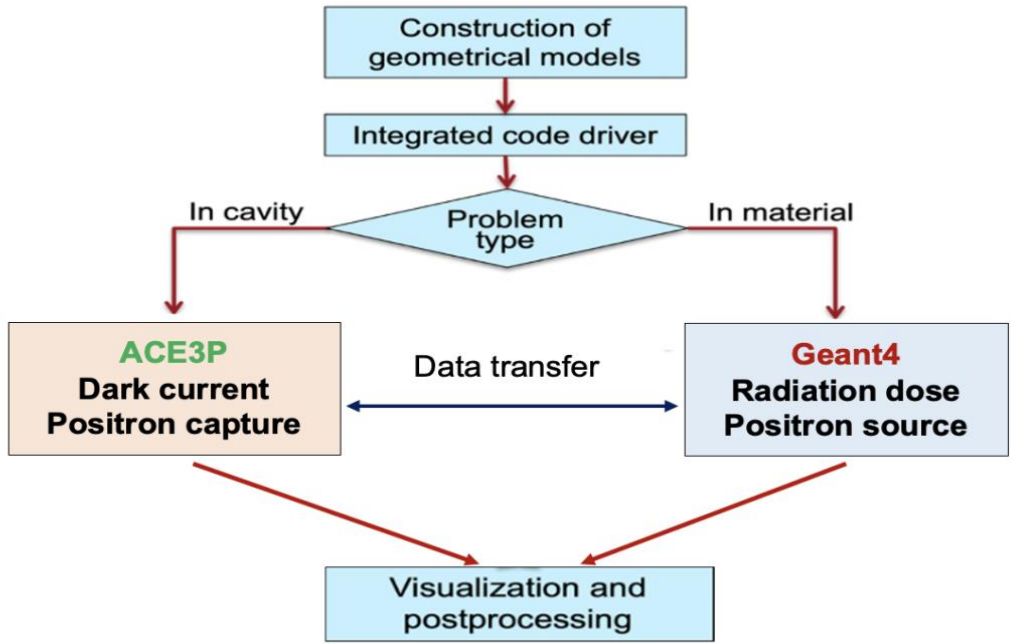

Figure 1. Workflow for integration of ACE3P and Geant4 simulation

**2.2 ACE3P and Geant4**

ACE3P is a comprehensive set of parallel multi-physics codes with electromagnetics, thermal and mechanical simulation capabilities developed at SLAC for almost two decades through the support of DOE Computational Grand Challenge and SciDAC [11] initiatives. It is based on high order curved finite elements for high-fidelity modeling and has been implemented on massively parallel computers for increased problem size and speed. All modules are highly parallelized running on high performance computing (HPC) platforms with thousands or more cores such as those at NERSC [12]. The use of high-order finite elements on tetrahedral conformal meshes with quadratic surfaces enables accurate and fast solution. ACE3P has been well accepted by the accelerator community as a benchmark and guidance of structure optimizations from large-scale simulations. Its eigensolver module Omega3P and S-parameters calculation module S3P are used widely in RF structure simulations, and its particle tracking module Track3P models multipacting and dark current effects in the structure. In this integrated tool, Omega3P/S3P will be used for EM field calculation, and Track3P for DC simulation.

Geant4 is a software toolkit for simulation of elementary particles passing through and interacting with matter. Geant4 is developed and maintained by the Geant4 Collaboration. Nowadays all LHC experiments at CERN, many current and near future experiments of DOE laboratories all rely on Geant4. Recently, a Geant4 based positron beam source package (GPos) has been developed at LBNL [13]. It is an easy-to-use publicly available C++ code, with support for hybrid MPI and openPMD [14] as well as parallel I/O tool created to model relativistic particle beam and solid target interactions.

Our goal is to develop a standalone software package for modeling particle-matter interactions radiation effects in accelerator cavity by integrating ACE3P for modeling of accelerator structures and Geant4 for calculation of interactions of particles with matter.

### 2.3 Data transfer between ACE3P and Geant4

A C++ API for openPMD [15] has been developed in ACE3P to convert ACE3P unstructured EM field data in NetCDF format from finite element discretization to structured data for Cartesian grid simulation used in other code, such as beam dynamics code IMPACT [16-18] and Geant4. The openPMD-API can execute in parallel using multiple compute cores with MPI, allowing faster data output. The electric field of a resonant mode of a simple pillbox cavity using the openPMD format is shown in Fig. 2.

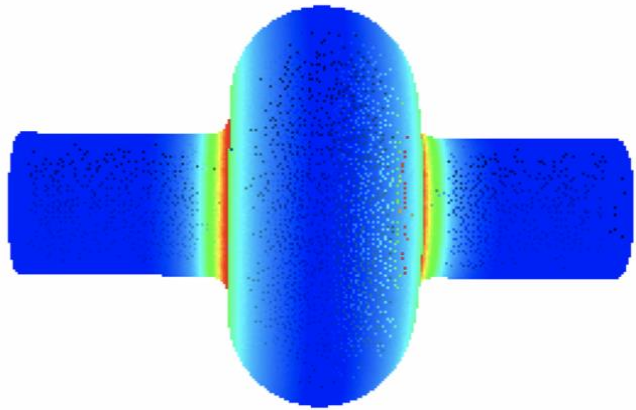

Figure 2. Visualization of electric field in openPMD format.

For particle data transfer from ACE3P to Geant4, an intermediary particle data reader/writer for particles 6D phase space data (**x**, **p**) and their timestamps (t) in ASCII format has been developed in both ACE3P and Geant4. We will expand the particle data transfer capability to include openPMD-API reader/writer to enhance I/O efficiency.

### 2.4 CAD Model

ACE3P and Geant4 can both start from a CAD model of a geometric domain.

The ACE3P modules, Omega3P, S3P and Track3P, discretize the CAD model into a finite element mesh consisting of curved finite elements and provide mesh representation knowledge of which finite element entities are on which boundaries of the CAD model. The way ACE3P handles CAD geometry is through the third-party mesh generator Cubit [19], which can read in a CAD model to generate finite element meshes for ACE3P simulation.

Geant4 employs a faceted representation of the CAD model boundary. The build-in GDML writer and reader will be used for CAD model import and export in Geant4. A community-standard format that most finite element software accepts is the Standard for the Exchange of Product Data (STEP), which is recognizable by ACE3P, but not directly by Geant4. For model import to Geant4, a converter tool is needed to convert a CAD model to Geant4 GDML recognized format. There are several developed convert tools for Geant4 CAD model import [20-23]. In this integration, Cubit is used to convert CAD model to STL format, which can be recognizable by Geant4 through its built-in GDML writer and reader.

## 3 APPLICATION

### 3.1 Dark current and radiation simulation on KEK 7 cell structure

KEK has performed DC simulations for a smaller model with 7 cells instead of 56 cells using the commercial EM code CST due to lack of computing power and memories. We benchmarked ACE3P DC simulations with CST using the same 7-cell structure.

#### 3.1.1 EM modelling using ACE3P

S3P, the S-parameter module in ACE3P, was used to solve for the operating mode in the 7-cell structure. Track3P, the particle tracking code module in ACE3P, was then used to track the particles under the EM fields from S3P. The ACE3P results are compared with CST for a field gradient of 21.8 MV/m as shown in Fig. 3.

#### 3.1.2 Radiation modelling using Geant4

The simulations from Track3P and CST show that particles mainly emit from the cavity disks, where higher electric fields are. Some particles hit the cavity wall and interact with it, which can be studied using Geant4. The primary particles collected from Track3P, the 7-cell structure solid model with the two couplers and the 7-cell vacuum cavity are loaded into Geant4 for radiation study. Fig. 4 shows the preliminary radiation simulation setup, and further radiation study will be carried out for the 56-cell real structure.

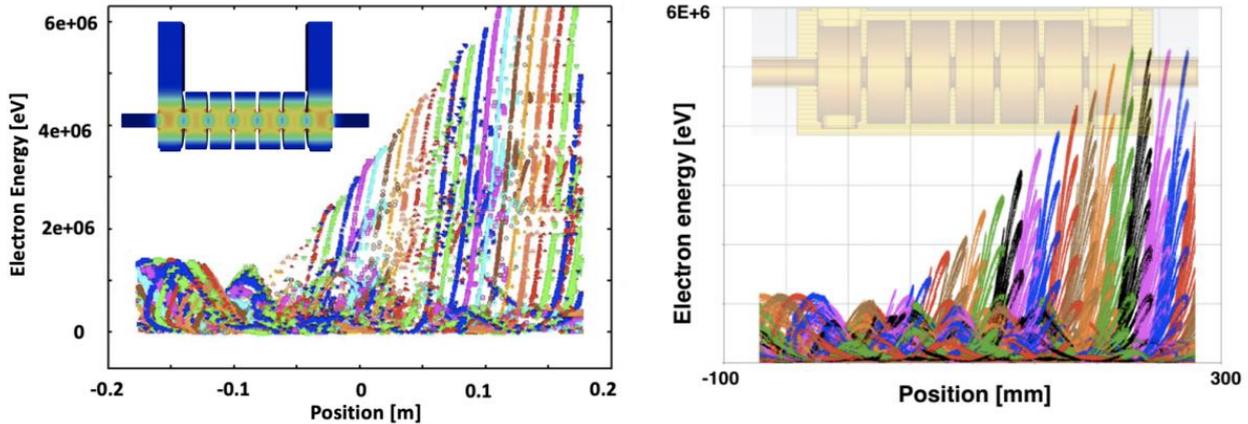

Figure 3: Energy spectrum from Track3P (left) and CST (right) in the 7-cell structure.

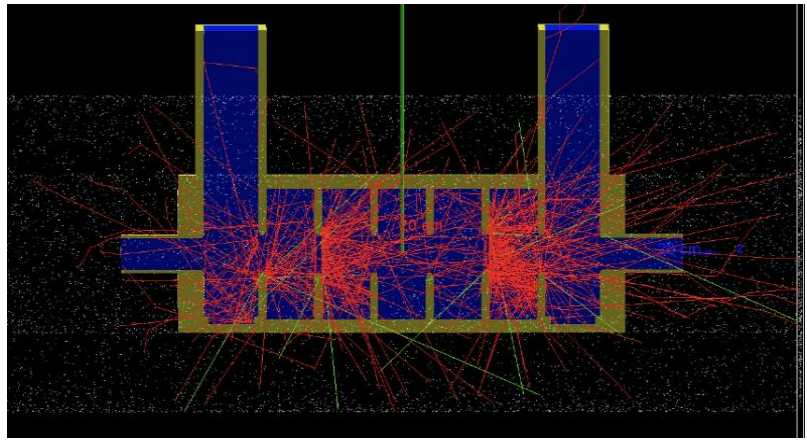

Figure 4. Particles trajectory on 7-cell structure by using Geant4.

### 3.2 Dark current and radiation simulation on KEK 56 cell structure

KEK performed dark current and radiated light intensity study on a 56-cell S-band accelerating structure [24]. Fig. 5 is the bench test of the S-band structure at KEK. Based on the S-band 56-cell model provided by KEK (Fig. 6), a preliminary dark current simulation is performed on NERSC Perlmutter supercomputer. The simulation procedure is as follows.

1) A mesh with 4M curved tetrahedral elements is generated for the full 56-cell vacuum region using Cubit, with denser mesh resolution around each cell iris region (Figure 7).

2) S3P, the S-parameter module in ACE3P, is used to solve for the operating mode with frequency at 2.856 GHz. It took several minutes to get 2nd order EM field by using 8 CPU nodes on NERSC Perlmutter. Fig. 7 shows the electric field magnitude.

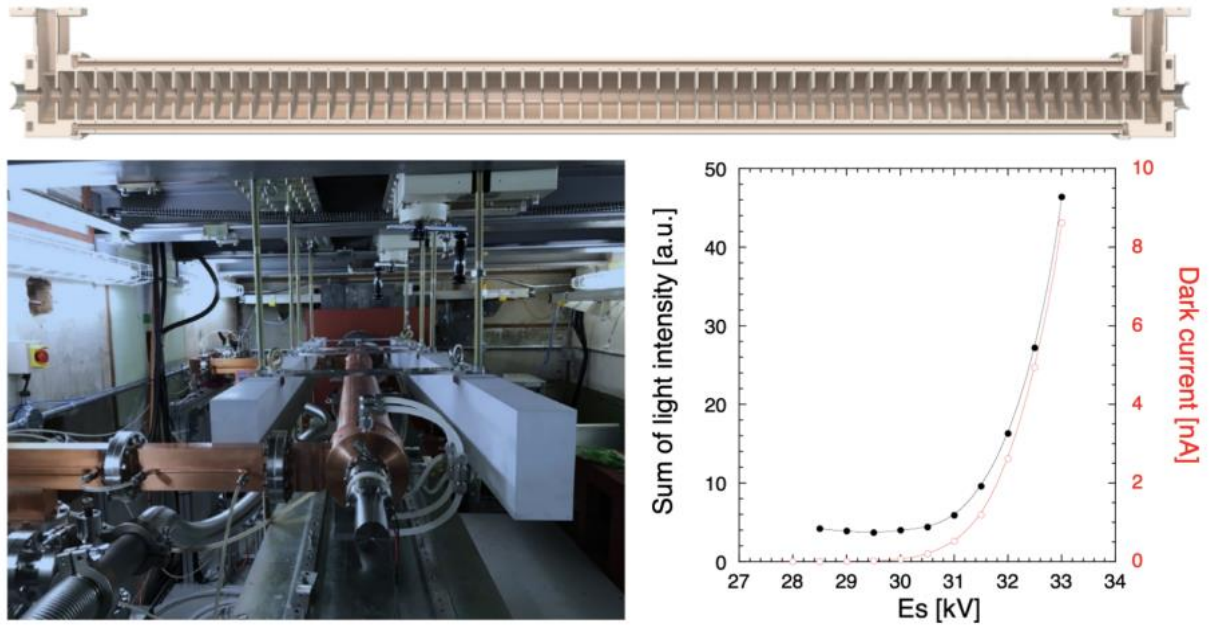

Figure 5. Bench test of an S-band structure at KEK.

3) Track3P, a particle tracking code in ACE3P, is used for dark current simulation. Electrons are emitted from the cavity surface according to the Fowler-Nordheim formula. The high-fidelity geometry representation built in the finite-element method allows for realistic modeling of particle emission on the cavity wall. These electrons contribute to the dark current, and their movements in the cavity are governed by the rf fields. When the electrons hit the cavity wall, they will be terminated in ACE3P simulation and their phase space information will be written to a file, which will be used for postprocessing and further study. For a typical dark current simulation, a total of 80k primary particles are emitted from the surface and it takes 25 RF cycles for particles transiting the whole 56 cells. The computational time was less than 30 minutes to complete the end-to-end simulation using 8 CPU Perlmutter nodes. Fig. 8 shows a snapshot of particle trajectories. This calculation will serve as a validation of the integrated simulation tool.

4) Dark current profiles observed on a phosphor screen would be used for validation of the integrated simulation code. Occasional high current spots have been detected on the screen and seem to cause extensive direct hitting and radiation damages to the structure and peripheral equipment. The exact simulation could trace the emission origins of the high dark current in the structure and play crucial role in radiation protection. The benchmark of dark current between ACE3P simulation and measured data from KEK is in progress.

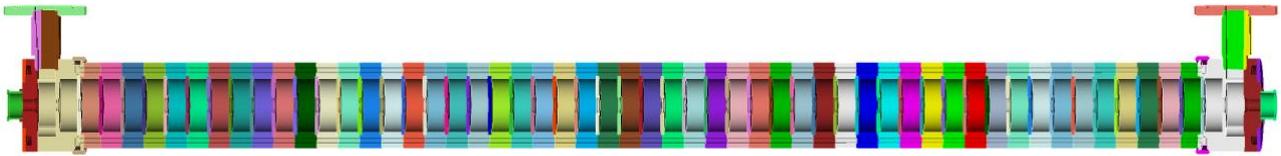

Figure 6. S-band 56-cell CAD model provided by KEK.

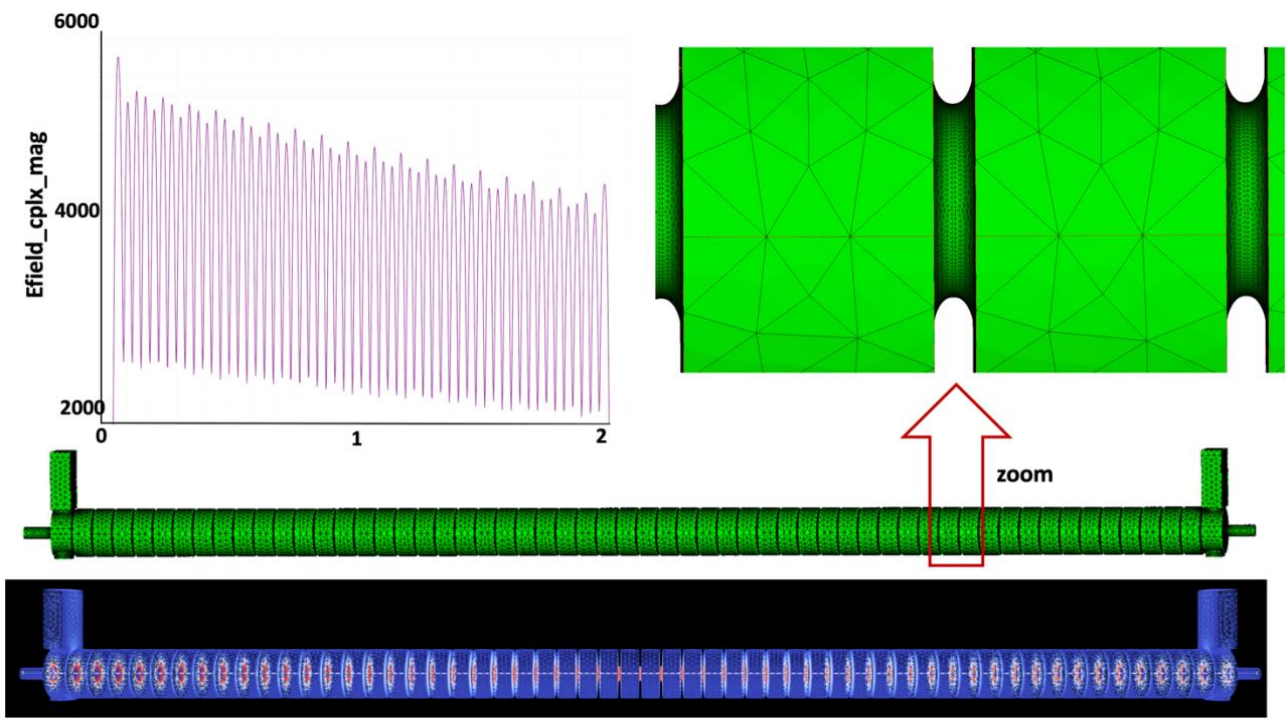

Figure 7. Mesh and complex E field magnitude profile for S-band 56-cell structure calculated by ACE3P.

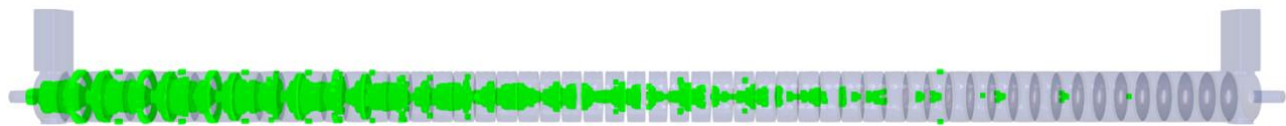

Figure 8. Snapshot of particle trajectories.

## 4  CONCLUSION AND FUTURE WORK

An integrated simulation tool for dark current radiation effects using ACE3P and Geant4 is under development. The reader/writer I/O for transferring of particle data from Track3P to Geant4 for DC radiation effects studies have been implemented. The development of geometry interface between ACE3P and Geant4 have been completed, enabling the use of CAD model in the integrated tool. A 7-cell structure dark current simulation have been benchmarked well with EM software CST. Dark current validation between ACE3P simulation and KEK measurement for 56-cell S-band accelerating structure is in progress.

In addition, there will be more capabilities development and more applications by using the developed tool, for example,

- Implement s Python script to facilitate simulation workflow on NERSC supercomputers.
- Perform thermal analysis using ACE3P from positron target simulation using Geant4.
- Complete positron generation simulation with diagnostics and visualization capabilities.

## Acknowledgments

This research used resources of the National Energy Research Scientific Computing (NERSC) Center, which is supported by the Office of Science of the U.S. Department of Energy under Contract No. DE-AC02-05CH11231.